\documentclass[aps,prl,twocolumn,showpacs,footinbib]{revtex4-1}

\usepackage[dvips]{graphicx}
\usepackage{amsmath}
\usepackage{subfigure}
\usepackage{psfrag}
\usepackage{color}

\newcommand{\nis}{\text{\it NIS}}
\newcommand{\nin}{\text{\it NIN}}

\begin{document}
\title{\bf Adiabatic quantum pumping in graphene NIS junctions}

\author{M. Alos-Palop and M. Blaauboer}

\affiliation{Kavli Institute of Nanoscience, Delft University of
  Technology, Lorentzweg 1, 2628 CJ Delft, The Netherlands}

\date{\today}

\begin{abstract}
  We investigate adiabatic quantum pumping through a normal
  metal-insulator-superconductor (NIS) junction in a monolayer of
  graphene. The pumped current is generated by periodic modulation of
  two gate voltages, applied to the insulating and superconducting
  regions respectively. In the bilinear response regime and in the
  limit of a thin high insulating barrier, we find that the presence
  of the superconductor enhances the pumped current per mode by a
  factor of 4 at resonance. Compared to the pumped current in an
  analogous semiconductor NIS junction, the resonances have a $\pi/2$
  phase difference. We also predict experimentally distinguishable
  differences between the pumped current and the tunneling conductance
  in graphene NIS junctions.
\end{abstract}

\pacs{72.80.Vp,74.45+c,73.23.-b}
\maketitle

Adiabatic pumping is a transport mechanism in meso- and nanoscale
devices by which a finite dc current is generated in the absence of an
applied bias by low-frequency periodic modulations of at least two
system parameters (typically gate voltages or magnetic
fields)~\cite{Buttiker1994,Brouwer1998}. In order for electrical
transport to be adiabatic, the period of the oscillatory driving
signals has to be much longer than the dwell time $\tau_{\rm dwell}$
of the electrons in the system, $ T = 2 \pi \omega^{-1} \gg \tau_{\rm
  dwell} $. Adiabatic {\it quantum} pumping~\cite{Spivak1995} refers
to pumping in open systems in which quantum-mechanical interference of
electron waves occurs. In the last decade, many different aspects of
quantum pumping have been investigated in a diverse range of
nanodevices, for example charge and spin pumping in quantum
dots~\cite{Switkes1999, *Mocciolo2002, *Sharma2003, *Watson2003}, the
relation of quantum pumping to geometric (Berry)
phases~\cite{Avron2000} and the role of electron-electron
interactions~\cite{[{See e.g.; }]Splettstoesser2005,
  *Sela2006,*Reckermann2010}. Quantum pumped currents have also been
studied in hybrid systems consisting of normal-metal (N) and
superconducting (S) parts, such as NS and SNS
junctions~\cite{Wang2001, Blaauboer2002, Taddei2004, Governale2005,
  Splettstoesser2007}.  Recently, investigations of quantum pumping in
graphene mono- and bilayers have appeared~\cite{Prada2009, *Zhu2009,
  *Prada2010, *Wakker2010}.

In this Letter we investigate adiabatic charge pumping through a
normal metal-insulating-superconductor (NIS) junction in graphene. The
pumped current is generated by adiabatic variations of two gate
voltages $U_0(t)$ and $V_0(t)$ which change, respectively, the Fermi
level in the superconducting region and the height of the insulating
tunnel barrier. The central question we aim to answer is what the
effect of electron-hole (Andreev) reflection is on pumped charge
currents in graphene. Using the scattering matrix formalism, we
calculate the adiabatically pumped current at zero temperature in the
linear response regime, i.e., for small variations of the pumping
parameters $U_0$ and $V_0$, and we compare this with the pumped
current in the absence of the superconducting lead. Our main result is
that the presence of the superconducting lead enhances the pumped
current per mode by a factor of 4 and the total pumped current by a
factor of $3 \sqrt{2}/2$ at the resonant tunneling condition. Off
resonance, the pumped current is an order of magnitude smaller than
the analogous current in a semiconductor NIS junction. We also find
that whereas the conductance \textit{increases} with $U_0$ for thin
barriers, the pumped current \textit{decreases} with $U_0$. This
difference might be used to discriminate between conductance and
pumped currents in graphene NIS junctions. In the last part of the
Letter, we briefly comment on the pumped current in the so-called
specular reflection regime (where $\Delta_0 \geq E_F$ with $\Delta_0$
the superconductor gap and $E_F$ the Fermi energy)
\cite{Beenakker2006}.

\begin{figure}
  \centering
  \psfrag{Ef}{\begin{picture}(0,0)
      \put(3,4){\makebox(0,0){\textbf{$E_F$}}}
    \end{picture}}
  \psfrag{0}{\begin{picture}(0,0)
      \put(22,3){\makebox(0,0){$0$}}
    \end{picture}} 
  \psfrag{-d}{\begin{picture}(0,0)
      \put(22,3){\makebox(0,0){$-d$}}
    \end{picture}} 
  \psfrag{U1}{\begin{picture}(0,0)
      \put(6,2){\makebox(0,0){\textbf{$U_0$}}}
    \end{picture}}  
  \psfrag{U2}{\begin{picture}(0,0)
      \put(-1,-2){\makebox(0,0){\textbf{$U_0$}}}
    \end{picture}}
  \psfrag{V1}{\begin{picture}(0,0)
      \put(4,2){\makebox(0,0){\bf{$V_0$}}}
    \end{picture}}
  \psfrag{V2}{\begin{picture}(0,0)
      \put(2,0){\makebox(0,0){\bf{$V_0$}}}
    \end{picture}}
  \psfrag{D_0}{\begin{picture}(0,0)
      \put(6,2){\makebox(0,0){\textbf{$\Delta_0$}}}
    \end{picture}}
  \psfrag{y}{\begin{picture}(0,0)
      \put(-12,3){\makebox(0,0){$\vec{y}$}}
    \end{picture}}
  \psfrag{x}{\begin{picture}(0,0)
      \put(0,0){\makebox(0,0){$\vec{x}$}}
    \end{picture}} 
  \psfrag{A}{\begin{picture}(0,0)
      \put(0,0){\makebox(0,0){\tiny $\vec{A}$}}
    \end{picture}}
  \psfrag{V}{\begin{picture}(0,0)
      \put(4,-4){\makebox(0,0){$V_0$}}
    \end{picture}}  
  \psfrag{U}{\begin{picture}(0,0)
     \put(2,-2){\makebox(0,0){$U_0$}}
    \end{picture}}
  \includegraphics[scale=.9]{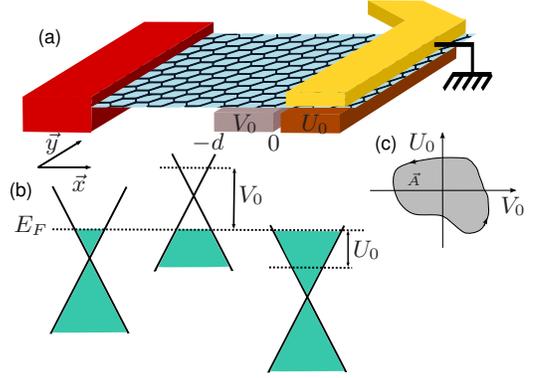}
  \caption[ss]{(a) Sketch of the graphene NIS junction. The variable
    gate voltage $V_0$ creates the insulating barrier I of length $d$
    and the gate voltage $U_0$ is applied to the superconducting
    electrode (yellow). (b) Schematic of the energy levels in the
    three regions. (c) The area enclosed in the $(U_0,V_0)$ parameter
    space during one pumping cycle.}
  \label{fig:model}
\end{figure}
Consider the geometry depicted in Fig.~\ref{fig:model}. A ballistic
sheet of graphene in the $(x,y)$-plane contains a potential barrier of
height $V_0$ and length $d$ and a superconducting contact in the
region $x \geq 0$. The barrier can be implemented by employing the
electric field effect ~\cite{Novoselov2004, Katsnelson2006} via a gate
voltage and superconductivity can be induced in the region $x\geq 0$
via the proximity effect. We assume the potential step $V_0$ to be
abrupt on both sides, which is justified close to the Dirac point
where the Fermi wavelength $\lambda_F \gg d$ and which can be realized
experimentally~\cite{Katsnelson2006}. The conductance $G(eV)$ through
such a graphene NIS junction has recently been studied, both for
potential barriers of finite length~\cite{Bhattacharjee2007} and in
the limit of a thin barrier~\cite{Bhattacharjee2006}. Here we
investigate the adiabatically pumped current through the latter
junction.  The pumped current is induced by periodic variations of
$U_0(t) = U_0 + \delta U_0 \cos (\omega t)$ and $V_0(t) = V_0 + \delta
V_0 \cos (\omega t + \phi)$.  The total pumped current $I$ into the
normal lead (the left contact in Fig.~\ref{fig:model}) can be
expressed as an integral over the area $A$ that is enclosed in
$(U_0,V_0)$ parameter space during one period, and is given by the
scattering matrix expression~\cite{Blaauboer2002}
\begin{eqnarray}
I \equiv I_N & = & \frac{\omega e}{2 \pi^2} \int_A\, dU_0\, dV_0\,
\sum_{\alpha, \beta \in N} \ \Pi_{\alpha\beta} (U_0,V_0)
\label{eq:Ip} \\
& \approx & \frac{\omega e}{2 \pi}\, \delta U_0\, \delta V_0\, \sin
\phi \, \sum_{\alpha, \beta \in N} \ \Pi_{\alpha \beta} (U_0,V_0)
\label{eq:IpLinearResp}
\end{eqnarray} 
with 
\begin{equation} 
\Pi_{\alpha \beta} (U_0,V_0) \equiv \mbox{\rm Im}\ \left(
  \frac{\partial S_{\alpha \beta}^{ee \star}}{\partial U_0}
  \frac{\partial S_{\alpha \beta}^{ee}}{\partial V_0} - \frac{\partial
    S_{\alpha \beta}^{he \star}}{\partial U_0} \frac{\partial
    S_{\alpha \beta}^{he}}{\partial V_0} \right).  
\end{equation}
Eq.~(\ref{eq:IpLinearResp}) is valid in the bilinear response regime
where $\delta U_0 \ll U_0$ and $\delta V_0 \ll V_0$ and the integral
in Eq.~(\ref{eq:Ip}) becomes independent of the pumping contour. The
indices $\alpha$ and $\beta$ sum over all modes in the normal lead and
$S$ denotes the Landauer-B\"uttiker scattering matrix whose elements
$S^{he}_{\alpha \beta, nm}$ describe the scattering of an electron in
mode $m$ in lead $\beta$ to a hole in mode $n$ in lead $\alpha$.

The low-energy excitations in the NIS junction close to the Dirac
point $K(K')$ are described by the $4\times 4$
Dirac-Bogoliubov-deGennes Hamiltonian~\cite{Beenakker2006}
\begin{equation} {\cal H} = \left( \begin{array}{cc}
  {\cal H}_a - E_F + U(x) & \Delta(x) \\
  \Delta^{*}(x) & E_F - U(x) - {\cal H}_a \end{array} \right)
\label{DBdG}
\end{equation}
with ${\cal H}_a = -i \hbar v_F (\sigma_x \partial_x + sgn(a) \sigma_y
\partial_y) $, where $sgn(a)$ is $\pm$ for $a = K(K')$, $v_F$ denotes
the Fermi velocity of the quasi-particles, and the potential $U(x)=
-U_0\theta(x) + V_0\theta(-x)\theta(x + d)$. The pair potential
$\Delta(x)$ which couples the electron and hole excitations has the
form $\Delta_0 e^{i \phi}$ and we assume that $(E_F + U_0) \gg
\Delta_0$, the mean-field condition for superconductivity. Two regimes
can be distinguished: $E_F \gg \Delta_0$ where the usual retro Andreev
reflection dominates, and $E_F \leq \Delta_0$ so $U_0 \gg \Delta_0$,
where specular Andreev reflection dominates \cite{Beenakker2006}.  Our
analysis focuses on the retro reflection regime unless we explicitly
mention that we study the specular reflection regime. The Hamiltonian
(\ref{DBdG}) acts on the four-component wavefunction $\psi_a
=(\psi_{Aa}, \psi_{Ba}, \psi^*_{A\bar{a}}, -\psi^*_{B\bar{a}})$, where
$A$ and $B$ denote the two nonequivalent sides of the graphene unit
lattice and $\bar{a} = K'(K)$ for $a = K(K')$.  Following
Ref.~\cite{Bhattacharjee2006}, we also introduce the dimensionless
barrier strength $\chi = d V_0 / (\hbar v_F)$ which allows us to
consider the limit of a thin barrier where $V_0 \rightarrow \infty$
and $d \rightarrow 0$ such that $\chi$ remains finite.

After applying continuity of the wavefunctions at the boundaries
$x=-d$ and $x=0$, one obtains the reflection and transmission
coefficients of the NIS junction, see Eqns.~(9) in
Ref.~\cite{Bhattacharjee2006}. At the Dirac point $\epsilon=0$, where
$\epsilon$ denotes the energy of the electrons measured from the Fermi
level \footnote{We remark that $\epsilon=0$ is the point of highest
  interest for observing quantum pumping, since it corresponds to the
  situation in which no bias voltage is applied. }, and defining
$\delta = E_F/(E_F + U_0)$, the derivatives of the coefficients for
normal and Andreev reflection, $r$ and $r_A$, with respect to the gate
voltages $U_0$ and $V_0$ are given by $\partial r_{(A)} / \partial U_0
=(\partial r_{(A)} / \partial \delta )(\partial \delta / \partial
U_0)$ and $\partial r_{(A)} / \partial V_0 =(\partial r_{(A)}
/ \partial \chi ) (\partial \chi / \partial V_0 )$ with
\begin{eqnarray}
\frac{ \partial r}{\partial \chi} &=& \frac{2 e^{i \alpha} \sin
    \gamma \cos \alpha (\cos 2 \chi + \sin \gamma \sin \alpha - i \cos
    \alpha \sin 2 \chi)}{(1 + \sin \gamma \sin \alpha \cos 2 \chi)^2},
  \nonumber \\ 
\frac{ \partial r^*}{\partial \delta} &=& \frac{-i e^{-i \alpha} \sin
  \alpha \cos \alpha (\cos 2 \chi \cos \alpha + i \sin 2 \chi)}{(1
    + \sin \gamma \sin \alpha \cos 2 \chi)^2}, \nonumber \\
\frac{ \partial r_A}{\partial \chi} &=& \frac{- 2 i e^{-i \phi} \sin
  \gamma \cos \gamma \sin \alpha \cos \alpha \sin 2 \chi}{(1 + \sin
  \gamma \sin \alpha \cos 2 \chi)^2} ,\nonumber \\
\frac{ \partial r_A^*}{\partial \delta} &=& \frac{- i 
  e^{i \phi} \cos \alpha
  \sin \alpha (\sin \gamma + \sin \alpha \cos 2 \chi)}
{\cos \gamma (1 + \sin \gamma \sin \alpha \cos 2 \chi)^2}.
\label{eq:derivativeCoefficients}
\end{eqnarray}
Here $\sin \gamma \equiv \delta \sin \alpha$ with $\alpha$ the angle
of incidence of the electron. Substituting
Eq.~(\ref{eq:derivativeCoefficients}) into Eq.~(\ref{eq:IpLinearResp})
and integrating over the angle of incidence yields the pumped current
at $\epsilon =0$
\begin{eqnarray}
  I^{\nis}_{g} &=&I_{g,0} \frac{\delta^2}{ \pi} \int_0^{\pi/2} d \alpha
  \frac{ \sin \gamma \sin \alpha \cos^4 \alpha}{(1 + \sin \gamma \sin
    \alpha \cos 2 \chi)^3}
\label{eq:IpE=0alphadep} 
\\ 
&\stackrel{U_0=0}{=}&I_{g,0} \frac{4 \cos^4
  \chi - 12 \cos^2 \chi + 8 \sqrt{2} |\cos \chi| - 3} 
{16\sqrt{2} |\cos \chi|(2 \cos^2 \chi -  1)^3 }
\label{eq:IpE=0}
\end{eqnarray}
where $I_{g,0} \equiv \omega e \frac{d}{\hbar v_F E_F} \delta U_0\,
\delta V_0\, \sin \phi$. Eq.~(\ref{eq:IpE=0}) is valid for $U_0
=0$. Note from Eq.~(\ref{eq:IpE=0alphadep}) that there is no
contribution to $I_g^{\nis}$ from normally incident electrons with
$\alpha=0$. This is a display of the Klein tunneling effect where the
Andreev reflection coefficient $|r_A(\alpha=0)|=1$ independent of
$\chi$ and $U_0$.

\begin{figure}
  \psfrag{x7}{\begin{picture}(0,0) \put(3,2){\makebox(0,0){\tiny
          $\chi$}}
  \end{picture}}
  \psfrag{pi/2}{\begin{picture}(0,0) \put(5,2){\makebox(0,0){\tiny
          $\frac{\pi}{2}$}}
  \end{picture} }
    \psfrag{pi}{\begin{picture}(0,0) \put(3,2){\makebox(0,0){\tiny
            $\pi$}}
  \end{picture}}
  \psfrag{pi2}{\begin{picture}(0,0) \put(6,2){\makebox(0,0){\tiny
          $\frac{3\pi}{2}$}}
  \end{picture}} 
 \psfrag{2pi}{\begin{picture}(0,0) \put(2,2){\makebox(0,0){\tiny
         2$\pi$}}
  \end{picture}} 
 \includegraphics[width=0.48\textwidth]{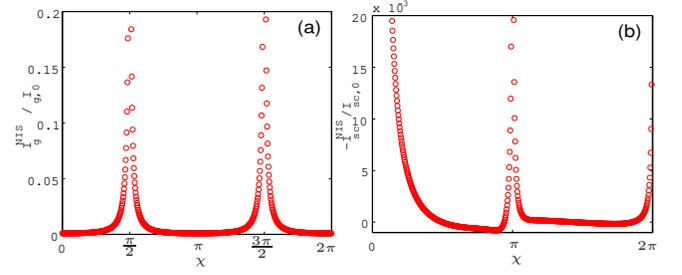}
 \caption{(a) The pumped current $I^{\nis}_g$ (Eq.~(\ref{eq:IpE=0}))
   as a function of the potential barrier strength $\chi$ in a
   graphene (g) NIS junction. (b) The analogous pumped current
   $I^{\nis}_{sc}$ (Eq.~(\ref{eq:IpSCalphadep})) in a semiconductor
   (sc) NIS junction, where $\sqrt{2 m E_F}d/\hbar=10$.}
\label{fig:Fig1Fig2}
\end{figure}
Figure~\ref{fig:Fig1Fig2}(a) shows the pumped current,
Eq.~(\ref{eq:IpE=0}), as a function of $\chi$. Notice that
$I_g^{\nis}$, just as the conductance in this system
\cite{Bhattacharjee2006}, is periodic in $\chi$ with a period
$\pi$. We see that $I_g^{\nis}$ reaches maximum values when $V_0
d=(n+1/2)\pi \hbar v_F $, where $n$ is an integer. This is the
condition for resonant transmission (i.e., $r=0$) when the conductance
of the system reaches a sharply peaked maximum
\cite{Bhattacharjee2006}. Due to the sharp changes in the derivative
of $r$ the pumped current diverges at this point.  Expanding
Eq.~(\ref{eq:IpE=0}) with respect to $\chi$ around $\chi=0$, we find
that $I_g^{\nis}$ scales as
\begin{eqnarray}
 I_g^{\nis} / I_{g,0} \sim a_0 + a_1 \; \chi^2
\label{eq:Ipscalingchi}
\end{eqnarray}
where $a_0 = \frac{16-11\sqrt{2}}{32} \approx 0.014$ and $a_1 = 3 (1 -
\frac{ 45}{32 \sqrt{2}})\approx 0.017$. The pumped current thus
increases with the barrier height close to $\chi=0$, which is another
manifestation of the Klein paradox. Next, we analyze the behavior of
the current with respect to $U_0$. We expand
Eq.~(\ref{eq:IpE=0alphadep}) with respect to $U_0$ around $U_0 = 0$ at
$\chi=0$, which results in
\begin{eqnarray}
 I_g^{\nis} / I_{g,0}  \sim  b_0 - b_1 \; U_0  
\label{eq:IpscalingU}
\end{eqnarray}
where $b_0 =a_0$ and $b_1 = \frac{3\sqrt{2}}{128}\approx
0.033$. Eq.~(\ref{eq:IpscalingU}) shows that the pumped current
decreases with increasing $U_0$. 

Figure \ref{fig:Fig1Fig2}(b) shows the corresponding pumped current
$I^{\nis}_{sc}$ in a {\it semiconductor} NIS junction. We model the
insulating region again as a barrier of height $V_0$ and width $d$,
and define the dimensionless barrier strength $\chi= \sqrt{2 m
  V_0}d/\hbar$. Solving the Bogoliubov-deGennes equation, matching the
wavefunction and its derivative at the boundaries, obtaining $r$ and
$r_A$ and calculating the derivatives with respect to the gate
voltages $V_0$ and $U_0$ yields the pumped current (at the Dirac point
$\epsilon=0$ and for $U_0=0$);
\begin{eqnarray}
  I_{sc}^{\nis} &=& I_{sc,0} \int_0^{\pi/2}
  d\alpha \cos \alpha \;k^3 d^3 \chi^3  
\label{eq:IpSCalphadep} \\
  &&
  \frac{\left( 2\chi (k^2 d^2 -\chi^2) - 
      (k^2 d^2 +\chi^2) \sin 2\chi \right)} 
  {( 2 k^2 d^2 \chi^2 \cos^2 \chi + (k^4 d^4 +\chi^4)\sin^2 \chi )^3}, 
\nonumber
 \end{eqnarray}
 where $I_{sc,0} \equiv  (\omega e / \pi) (2m^2 d^4 /\hbar^4) \delta
 U_0 \delta V_0 \sin \phi$ and $k\equiv(\sqrt{2 m E_F}/ \hbar )\cos
 \alpha$. Eqns.~(\ref{eq:IpE=0}) and (\ref{eq:IpSCalphadep}) are the
 main results of this Letter. From Eq.~(\ref{eq:IpSCalphadep}) we
 notice that, in contrast to graphene, the normally incident electrons
 in the semiconductor junction do contribute to the pumped current,
 illustrating the absence of Klein tunneling in a semiconductor NIS
 junction. Figure~\ref{fig:Fig1Fig2}(b) displays $I_{sc}^{\nis}$ as a
 function of the barrier strength $\chi$. We observe that
 $I_{sc}^{\nis}$ also oscillates as a function of $\chi$ with a period
 of $\pi$. However, the maxima at resonant transmission occur when
 $\chi = n \pi$ and are thus shifted by $\pi/2$ with respect to the
 maxima of $I_{g}^{\nis}$. Notice that $I_{sc}^{\nis}$ and
 $I_{g}^{\nis}$ mostly have opposite signs and that $I_{sc}^{\nis}$
 switches sign at several points. In addition, the pumped current in a
 semiconductor NIS junction is roughly one order of magnitude larger
 than in graphene \footnote{Here we used that
   $I_{sc}^{\nis}=I_{g}^{\nis} (2 m E_F)^{3/2}d^3/(\pi \hbar^3)$.}.

 An important question arising when thinking about experimental
 detection of pumped currents is how to distinguish them from the
 conductance $G$ in the system. In order to answer this question we
 explicitly compare both quantities. The conductance of the NIS
 junction in graphene was considered in Ref.~\cite{Bhattacharjee2006}
 and is given by, at $\epsilon =0$ and $U_0=0$,
\begin{equation}
  G_g^{\nis}=G_0\frac{\sqrt{A}(A+3) + (A^2 - 2A-3) 
    \arctan{(\sqrt{A})}}{A^{5/2}},
\end{equation}
where $G_0 \equiv \frac{4e^2}{h}\frac{E_F w} {\pi \hbar v_F}$ with $w$
the width of the sample and $A \equiv\cos 2\chi$.  Expanding
$G_g^{\nis}$ for small $\chi$ around $\chi=0$ at $U_0=0$ and for small
$U_0$ around $U_0=0$ at $\chi=0$ yields, respectively,
\begin{eqnarray}
 G_g^{\nis} /G_0 &\sim& (4 -\pi) + (16-5\pi)\chi^2 ,
\label{eq:GscalingChi}\\
 G_g^{\nis} /G_0&\sim&  (4-\pi) +(2-\pi/2) U_0/E_F.
\label{eq:GscalingU}
\end{eqnarray}
We find that $G_g^{\nis}=(4-\pi)G_0$ for $U_0=\chi=0$, somewhat below
the ballistic value $G_0$ due to mismatch in Fermi wavelength in the
normal and superconducting leads, as mentioned earlier
\cite{Beenakker2006}. Comparing the scaling behavior of $G_g^{\nis}$
and $I_g^{\nis}$ as a function of $\chi$,
Eqns.~(\ref{eq:Ipscalingchi}) and (\ref{eq:GscalingChi}), we find that
both transport quantities increase with increasing $\chi$. However
when comparing Eqns.~(\ref{eq:IpscalingU}) and (\ref{eq:GscalingU}) we
see that whereas the pumped current $I_g^{\nis}$ {\it decreases} with
increasing $U_0$, the conductance $G_g^{\nis}$ {\it increases} with
increasing $U_0$. Intuitively, switching on $U_0$ increases the Fermi
level mismatch, which increases the conductance~\footnote{See also
  Ref.~\cite{Beenakker2006}, where it was assumed that $U_0 \gg
  E_F$. The conductance is then given by $G=(4/3)G_0$ at zero bias
  $eV=0$ and decreases to the value $G=(4-\pi)G_0$ for $eV\gg
  \Delta_0$, in which case the Fermi level mismatch is minimal.}
however decreases the pumped current. This difference can be used to
discriminate the pumped current from the conductance in an actual
experiment.

\begin{figure}
  \psfrag{c}{\begin{picture}(0,0)
      \put(0,0){\makebox(0,0){\tiny \bf{$\chi$}}}
  \end{picture}}   
  \psfrag{p1}{\begin{picture}(0,0)
      \put(0,0){\makebox(0,0){\tiny\bf{$\pi$}}}
  \end{picture}}   
  \psfrag{p2}{\begin{picture}(0,0)
      \put(0,0){\makebox(0,0){\tiny $\pi$/2}}
  \end{picture}}   
\psfrag{0}{\begin{picture}(0,0)
      \put(0,2){\makebox(0,0){\tiny $0$}}
  \end{picture}} 
\includegraphics[width=0.48\textwidth]{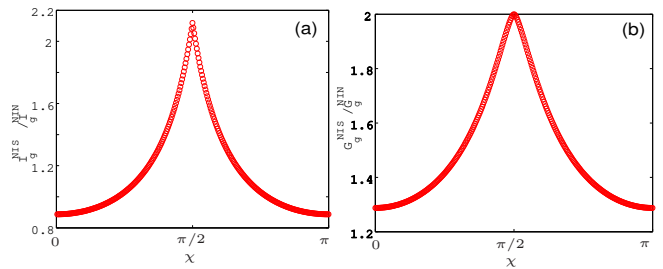}
\caption{Ratio of the pumped current (left) and the conductance
  (right) of the NIS junction and the NIN junction in graphene versus
  $\chi$ for $\epsilon= 0$ and $U_0=0$.}
\label{fig:Fig4Fig5}
\end{figure}
We now investigate the influence of the superconducting lead on the
pumped current by comparing $I_g^{\nis}$ with the pumped current
$I_g^{\nin}$ through an entirely normal NIN junction in graphene. At
the Dirac point $\epsilon =0$ and for $U_0=0$ the latter is given by
\begin{equation}
  I_g^{\nin}= I_{g,0} \frac{(1-|\cos \chi|)^2}{16 |\cos \chi| \sin^4 \chi}.
\label{eq:IpNIN}
\end{equation}
Similarly, the conductance through a NIN junction is given by
\begin{equation}
 G_g^{\nin}= G_0 \frac{\sin \chi - \cos^2 \chi \;
   \text{arctanh}(\sin \chi)}{\sin^3 \chi} .
\label{eq:G_NIN}
\end{equation}
Figure~\ref{fig:Fig4Fig5} shows the ratios $I_g^{\nis}/I_g^{\nin}$ and
$G_g^{\nis}/G_g^{\nin}$ versus $\chi$. In both cases, the
superconducting lead enhances the transport reaching a maximum at
$\chi=\pi/2$. For the conductance this maximum enhancement is 2 due to
the contribution of the holes \cite{Beenakker1994}, while the
enhancement of the pumped current rises from $2(16-11\sqrt{2})
\approx0.89 $ at $\chi=0$ to a factor of $3\sqrt{2}/2\approx 2.12$ at
$\chi=\pi/2$. However, when comparing $I_g^{\nis}$
(Eq.~(\ref{eq:IpE=0alphadep})) and $I_g^{\nin}$ per mode at
$\chi=\pi/2$, i.e., before integration over $\alpha$, we see that the
superconducting lead enhances the pumped current of each mode by a
factor of 4. This last result is due to both the holes which
contribute a factor of 2 and the asymmetry of the NIS junction with
respect to injection of charge carriers which contribute another
factor of~2~\cite{Blaauboer2002}.

At this point we briefly mention the behavior of the pumped current as
a function of an applied bias voltage~$eV$ both in the normal (retro)
and in the specular Andreev reflection regime \footnote{A more
  extensive analysis is in preparation and will be published
  elsewhere}. For bias voltages below the gap $eV \leq \Delta_0$, the
pumped current $I_g^{\nis}$ in the retro reflection regime decreases
from a finite value at $eV=0$ (see Eq.~(\ref{eq:IpE=0})) to zero at
$eV=\Delta_0$. At $eV=\Delta_0$, the particles are fully Andreev
reflected ($|r_A|=1$ independent of $\chi$ and $U_0$) and therefore
$I_g^{\nis} =0$. In the specular reflection regime
\cite{Beenakker2006} where $E_F \leq \Delta_0$ and $U_0 \gg E_F$, the
pumped current exhibits different behavior. First, the pumped current,
just as the conductance \cite{Bhattacharjee2006}, is insensitive to
$\chi$ for energies below the gap. The large mismatch in Fermi
energies of the normal and the superconducting leads already acts as a
barrier, and as a result the addition of another barrier is
irrelevant, explaining this behavior. Furthermore, the pumped current
$I_g^{\nis}$ is zero for bias voltages equal to the Fermi level $eV =
E_F$, due to the absence of Andreev reflection \cite{Beenakker2006},
and also for $eV= \Delta_0$, see the discussion above. As a final
remark, the pumped current is several orders of magnitudes smaller
than in the retro reflection regime.

Finally, we briefly comment on possibilities for experimental
observation of our predictions. Experiments with superconducting
electrodes on top of graphene in which multiple Andreev reflections
were observed have already been carried out
\cite{Heersche2007,Miao2007}. From these experiments we can estimate
the order of magnitude of the pumped current. Some typical parameters
are $\omega/(2\pi)=5 \;\text{GHz}$, $E_F=80\;meV$, $v_F =10^{6}\;m/s$
and barrier width $d = 10-20 \;nm$
\cite{Novoselov2005,*Zhang2005,*Novoselov2006,Katsnelson2006}. For
gate voltages on the order of $10\;meV$, the pumped current is on the
order of $10fA$ far from the resonant tunneling condition, going up to
$0.1-1\;pA$ or higher close to resonance.

In conclusion, we have investigated adiabatic quantum pumping in a
graphene NIS junction, which is generated by periodic modulation of
the insulating barrier $V_0$ and the Fermi level on the superconductor
side. We have demonstrated that the presence of the superconducting
lead can enhance the pumped current per mode by a factor of 4 (at
resonance) and suggested experimentally observable differences between
the conductance and the pumped current in this system.

\begin{acknowledgments}
This work has been supported by the Netherlands Organization for
Scientific Research (NWO/FOM).
\end{acknowledgments}

\bibliography{../general}
\end{document}